\documentclass{workshop}
\usepackage{natbib_mod}
\usepackage{amssymb}
\usepackage{amsmath}

\begin{document}

% TITLE OF THE PAPER
%  If the title is too long for a single line, you can split it 
%  by putting two backslashes. 
%  You might want to put the subtitle. Then it should be inserted 
%  within {\large  }.
%  e.g.:  
%     \title{ Too Long Title \\ for one line \\
%     {\large -- Subtitle --} }
\title{An X-ray spectral variability of fast disk winds in AGN}

% AUTHOR(S) 
\author{
Kouichi Hagino,$^1$ Hirokazu Odaka,$^2$ Chris Done,$^{1,3}$ Shin Watanabe,$^{1,4}$ Tadayuki Takahashi$^{1,4}$
\\[12pt]  % TO BE SPACED WITH ONE LINE
%
% INSTITUTES OF AUTHORS
$^1$  Institute of Space and Astronautical Science (ISAS), Japan Aerospace Exploration Agency (JAXA),\\
3-1-1 Yoshinodai, Chuo, Sagamihara, Kanagawa
252-5210, Japan\\
$^2$  KIPAC, Stanford University, 452 Lomita Mall, Stanford, CA 94305, USA\\
$^3$  Department of Physics, University of Durham, South Road, Durham DH1 3LE, UK\\
$^4$  Department of Physics, University of Tokyo, 7-3-1 Hongo, Bunkyo, Tokyo 113-0033, Japan\\
%
% please put the first author's initial and e-mail address below
\textit{E-mail(KH): hagino@astro.isas.jaxa.jp} 
%              \_ Initial    \
%                             \_ E-mail address
}

\abst{
Recent X-ray observations of blue-shifted absorption lines revealed an
existence of the extremely fast disk winds with outflow velocities of
$\sim0.1\textrm{--}0.3c$. Such fast outflows would have a large impact
on the coevolution of black holes and host galaxies since they are
expected to carry a large amount of kinetic energy. One of the common
characteristics of these fast winds is a strong time variability of the
absorption feature. To investigate this variability, we have developed a
new X-ray spectral model of the disk winds, which is generated by
3-dimensional Monte Carlo radiation transfer simulations on the
assumption of the realistic wind geometry. By applying our wind model
to the multi-epoch X-ray data of an archetypal wind source PDS~456,
we find the variability in the absorption line is explained by a change of
the wind outflowing angle without any large variability in the mass
outflow rate of the wind. This result indicates that the fast disk winds
are stable and that local hydrodynamic instabilities produce a large
time variability of the absorption line.
Moreover, we also apply our wind model to the disk-line source
1H~0707$-$495. The characteristic Fe-K feature in this source is
successfully reproduced our disk wind model for all the observations.
%up to 4 pages
}

\kword{black hole physics --- radiative transfer --- galaxies: individual (PDS 456, 1H 0707$-$495)}

\maketitle
\thispagestyle{empty}

\section{Introduction}
Blueshifted absorption lines with velocities of $0.1\textrm{--}0.3c$ from
highly ionized iron ions have been discovered in many local active
galactic nuclei (AGN) by recent X-ray observations
\citep[e.g.,][]{Chartas2002,Reeves2003,Pounds2003a,Pounds2003b}. These
absorption lines indicate that heavy ions are outflowing with such
extremely fast velocities from the vicinity of supermassive black holes
at the center of galaxies. These outflows are often referred to as
ultra-fast outflows (UFOs) \citep{Tombesi2010}. Since they are
expected to carry large amounts of kinetic energy to the host galaxies,
they are thought to play an important role in the co-evolution of the
black holes and the galaxies \citep{King2015}.

\begin{figure}[tbp]
\centering
\includegraphics[angle=180,width=0.9\hsize]{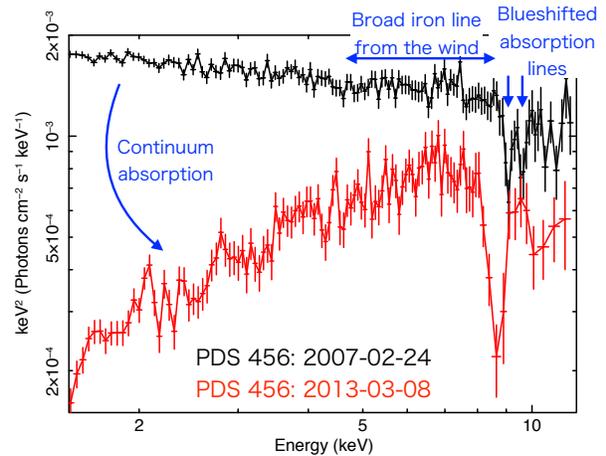}
\vspace{-0.4cm}
\caption{Observed spectra of an archetypal ultra-fast outflow in PDS~456.}
\label{fig:ufospec}
\end{figure}

\begin{figure*}[tbp]
\centering
\includegraphics[width=0.9\hsize]{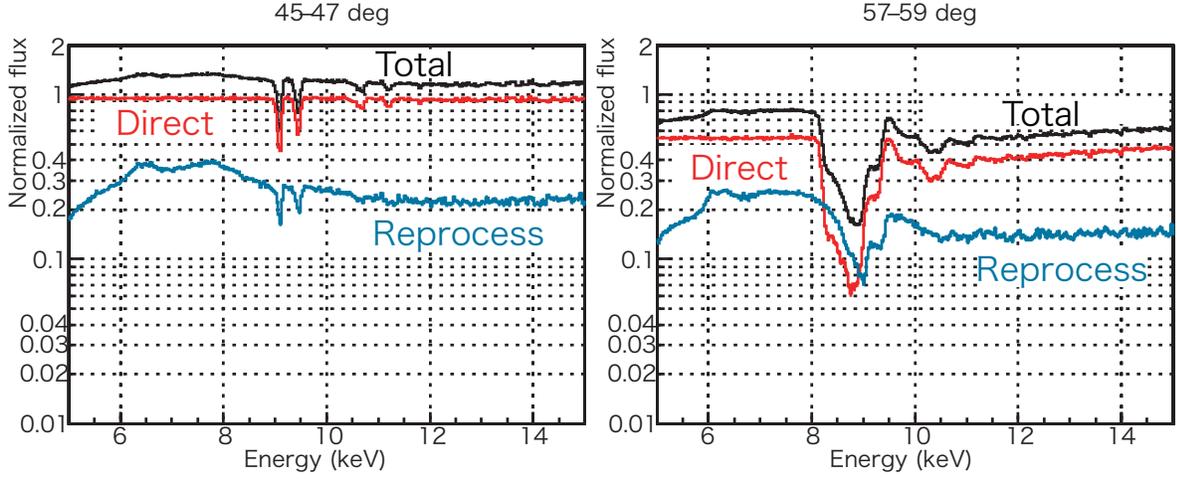}
\vspace{-0.5cm}
\caption{Simulated spectra with different angle \citep{Hagino2016}.}
\label{fig:simspec}
\end{figure*}

In spite of its importance, physical properties of UFOs are still unclear. Particularly, the origin of spectral variation of the blueshifted absorption lines in UFOs is unknown. As an example, spectra of PDS~456 observed in 2007 and 2013 by \textit{Suzaku} are shown in Fig.~\ref{fig:ufospec}. These spectra clearly show spectral changes in both absorption lines and continuum absorption, which are a common interesting characteristic in UFOs. Since the blueshifted absorption lines are usually interpreted as K-shell transitions of He-like and/or H-like iron ions, outflowing materials should be highly ionized with an ionization parameter of $\xi\equiv L/nR^2\sim10^{3\textrm{--}4}$. On the other hand, such highly ionized material is not able to explain the strong continuum absorption as shown in Fig.~\ref{fig:ufospec}, which requires relatively low ionized materials with $\xi\lsim10^2$. Thus, these materials are often thought to be cool clumps embedded in the hot highly ionized outflow. These clumps are naturally expected in hydrodynamic simulations of the accretion disk winds due to ionization instability and/or hydrodynamic instability \citep{Krolik1981,Takeuchi2014}. While occultation by the cool clumps can easily explain the spectral variability in continuum absorption \citep{Matzeu2016}, there is no clear picture for explaining the spectral variation in blueshifted absorption lines produced by the highly ionized outflow.

\section{Monte Carlo radiation transfer simulations}
We construct a new spectral model of the UFO in order to understand the spectral variability of the highly ionized absorption lines \citep{Hagino2015}. In this model, we assume the UV-line driving mechanism. This is the best candidate of acceleration mechanism of UFOs because of its good efficiency in AGN environment. In the UV-line driving mechanism, moderately ionized materials are accelerated by radiation pressure due to bound-bound transitions with UV photons. Since cross sections of bound-bound transitions are 3--4 orders of magnitude larger than that of Thomson scattering, this mechanism can efficiently accelerate and launch the fast disk wind if materials are appropriately ionized. Moreover, since AGN radiation typically peaks at the UV band, which moderately ionizes materials and causes radiation pressure via bound-bound transitions, AGN is suitable for the UV-line driving.

We assume a biconical geometry, which was developed for accretion disk winds in cataclysmic variables \citep{Knigge1995}. Based on the hydrodynamic simulations of the UV-line driven disk winds, a covering fraction of $\Omega/4\pi=0.15$ is adopted \citep{Proga2000, Nomura2016}. A maximum velocity of the disk wind is determined by the observed absorption lines, and its distribution in the disk wind is calculated following the CAK velocity law \citep{Castor1975}. A density distribution is derived from the velocity distribution and conservation of mass. Then, an ionization structure of the wind is calculated 1-dimensionally along stream lines by utilizing XSTAR \citep{Kallman2001}, which outputs ion abundances and electron temperatures in the wind. The radiation transfer calculation is performed in these density, velocity, ionization distributions.

We adopt Monte Carlo method for the radiation transfer calculation in the UV-line driven disk wind because its non-spherical geometry makes it very difficult to solve with analytical methods. The Monte Carlo radiation transfer simulation is performed by utilizing a simulation framework called \textsc{MONACO} \citep{Odaka2011}. \textsc{MONACO} is a general-purpose framework, which have been applied for synthesizing X-ray spectra from many kinds of astrophysical objects \citep[e.g.,][]{Odaka2014,Odaka2016,Furui2016,Hagino2015,Hagino2016}. In this code, all important physical processes in photoionized plasma are implemented, namely photoionization, recombination, photoexcitation, Compton scattering \citep{Watanabe2006}. In addition to these, special relativistic effects, which are essential in UFOs, are also considered.

\begin{figure}[tbp]
\centering
\includegraphics[width=0.9\hsize]{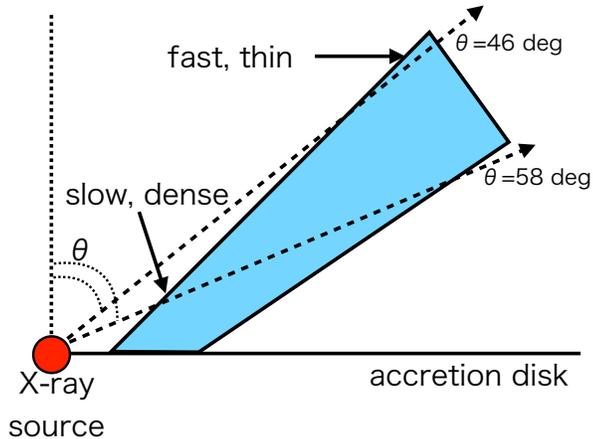}
\vspace{-0.2cm}
\caption{A schematic view of angle dependences of the absorption lines.}
\label{fig:angle}
\end{figure}

X-ray spectra calculated by our simulations are shown in Fig.~\ref{fig:simspec}. As shown in the left panel, blueshifted absorption lines and a broad emission line similar to the observations are reproduced. More interestingly, when the wind is observed from a larger viewing angle, the absorption lines become deeper and broader as shown in the right panel. Fig.~\ref{fig:angle} schematically show how the absorption line features depend on the observer's viewing angle. At a smaller viewing angle, only a thin and fast part of the wind is observed. On the other hand, at a larger angle, a slow and dense part is observed as well as the fast and thin part, and it makes the absorption lines broader and deeper.

\section{Application to an archetypal wind source PDS~456}
We apply our spectral model to observational data of an archetypal disk wind source PDS~456. This source is observed by \textit{Suzaku} for 5 times between 2007 and 2013, and most of these data clearly show blueshifted absorption lines \citep{Reeves2009,Gofford2014}. Also, as already shown in Fig.~\ref{fig:ufospec}, the absorption line features strongly vary. It means that this source is one of the best target to study the origin of the spectral variation in the highly ionized absorption lines.

\begin{figure*}[t]
\centering
\includegraphics[width=0.45\hsize]{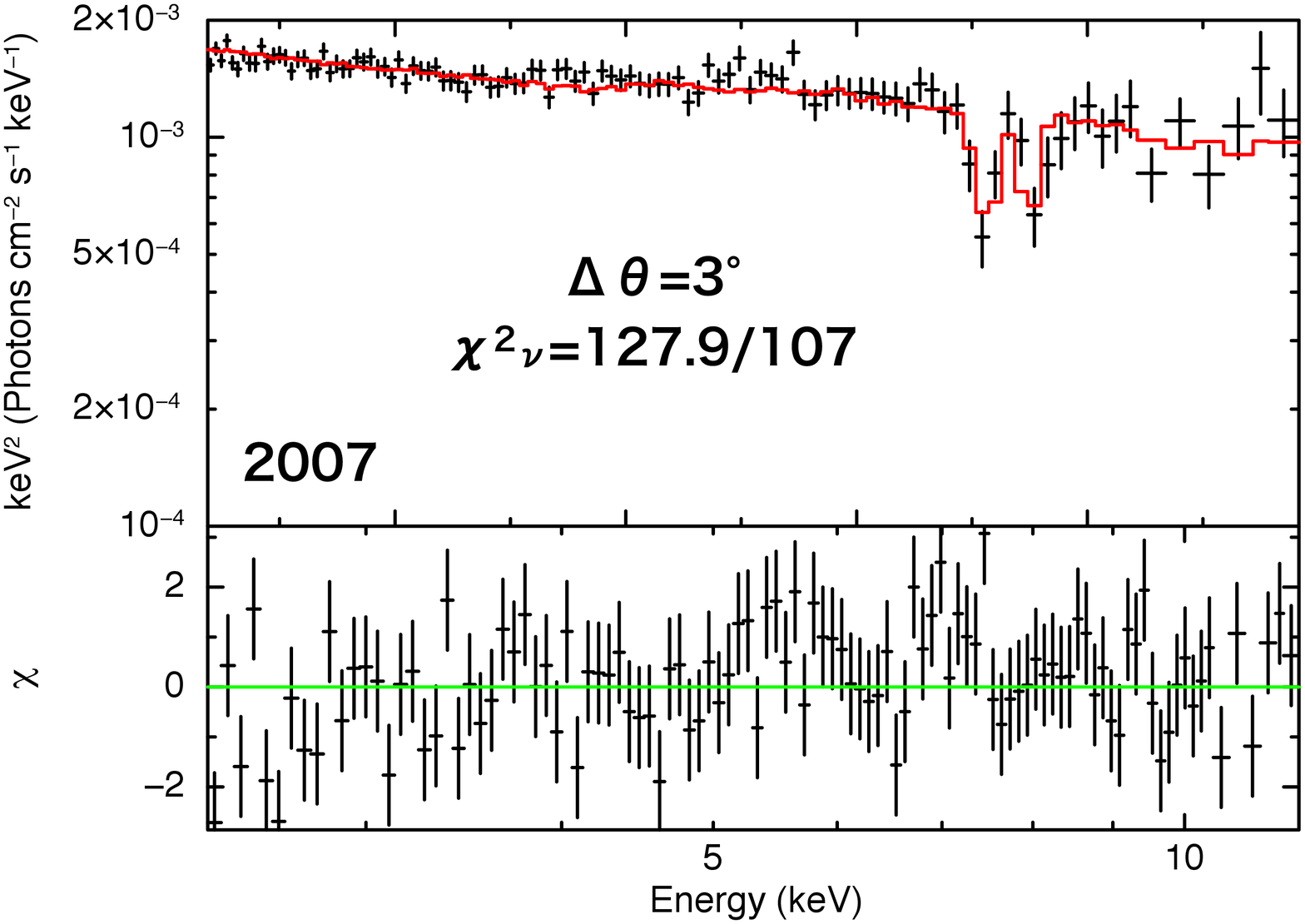}
\includegraphics[width=0.45\hsize]{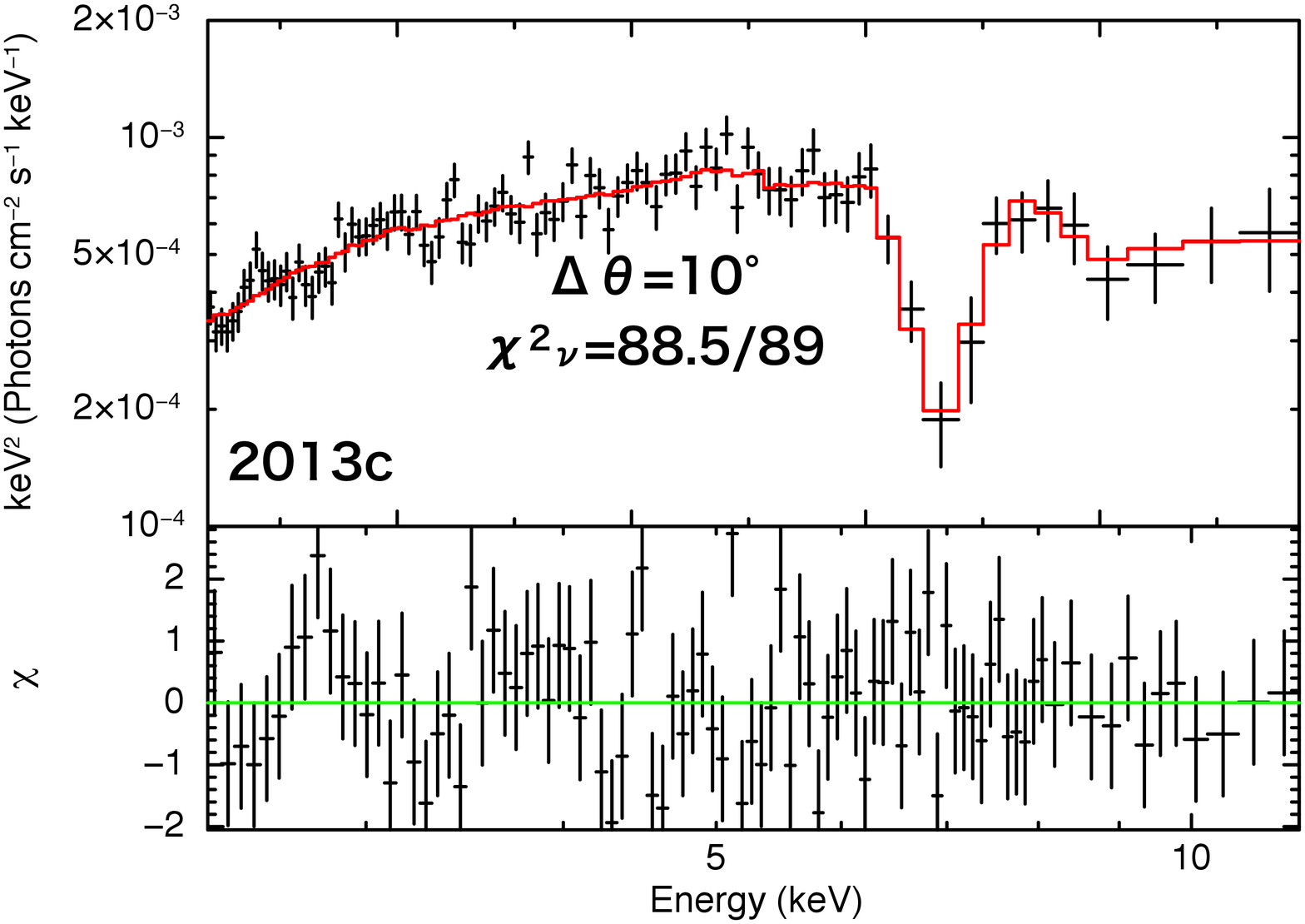}
\vspace{-0.2cm}
\caption{\textit{Suzaku} spectra of PDS~456 compared with our spectral model. $\Delta\theta$ indicates a difference between the observer's line of sight and an inner stream line of the wind \citep{Hagino2015}.}
\label{fig:pds}
\end{figure*}

All the \textit{Suzaku} spectra of PDS~456 are successfully reproduced by our wind model with a mass outflow rate of $\dot{M}_\mathrm{wind}/\dot{M}_\mathrm{Edd}=0.13$ and a terminal velocity of $v_\mathrm{wind}\sim0.3c$. As examples, we show the spectra and best-fit models for observations in 2007-02-24 and 2013-03-08 in Fig.~\ref{fig:pds}. Between the observations, we change only the viewing angle and the terminal velocity without changing the global parameters such as the mass outflow rate.

A change of the viewing angle, which successfully explains the strong spectral variability of absorption lines, possibly relate to a flapping or/and an inhomogeneity of the disk winds. They are naturally expected in hydrodynamic simulations of the winds due to hydrodynamic instability \citep{Proga2000,Krolik1981,Takeuchi2014}. Thus, our result indicates that the strong variability of the highly ionized absorption lines in UFOs could originate from a local instability of the wind, not from a global change of the wind.

\section{Application to a disk-line source 1H~0707$-$495}
As a next target, we choose a narrow line Seyfert 1 galaxy 1H~0707$-$495, which is famous for its extremely smeared disk-line feature \citep{Fabian2004}. However, this disk-line interpretation for a characteristic iron-K spectral feature in this source requires extreme conditions. Since the iron-K line is extremely smeared, a black hole spin must be close to maximum. Moreover, an X-ray emitting region (corona) must be located very close to the event horizon to explain its strong reflection component. Also, a large iron abundance of 7--20 is required by its strong iron line.

Alternatively, we interpret the iron-K feature as an absorption feature produced by the UFO. This interpretation is very natural because the spectral features in 1H~0707$-$495 is very similar to those in PDS~456. Also, this interpretation is supported by optical data of this source. It requires super Eddington accretion \citep{Done2016}, and in such a situation, the accretion disk must not be a standard thin disk, which is assumed in the disk-line interpretation.

\begin{figure*}[htbp]
\centering
\includegraphics[width=0.45\hsize]{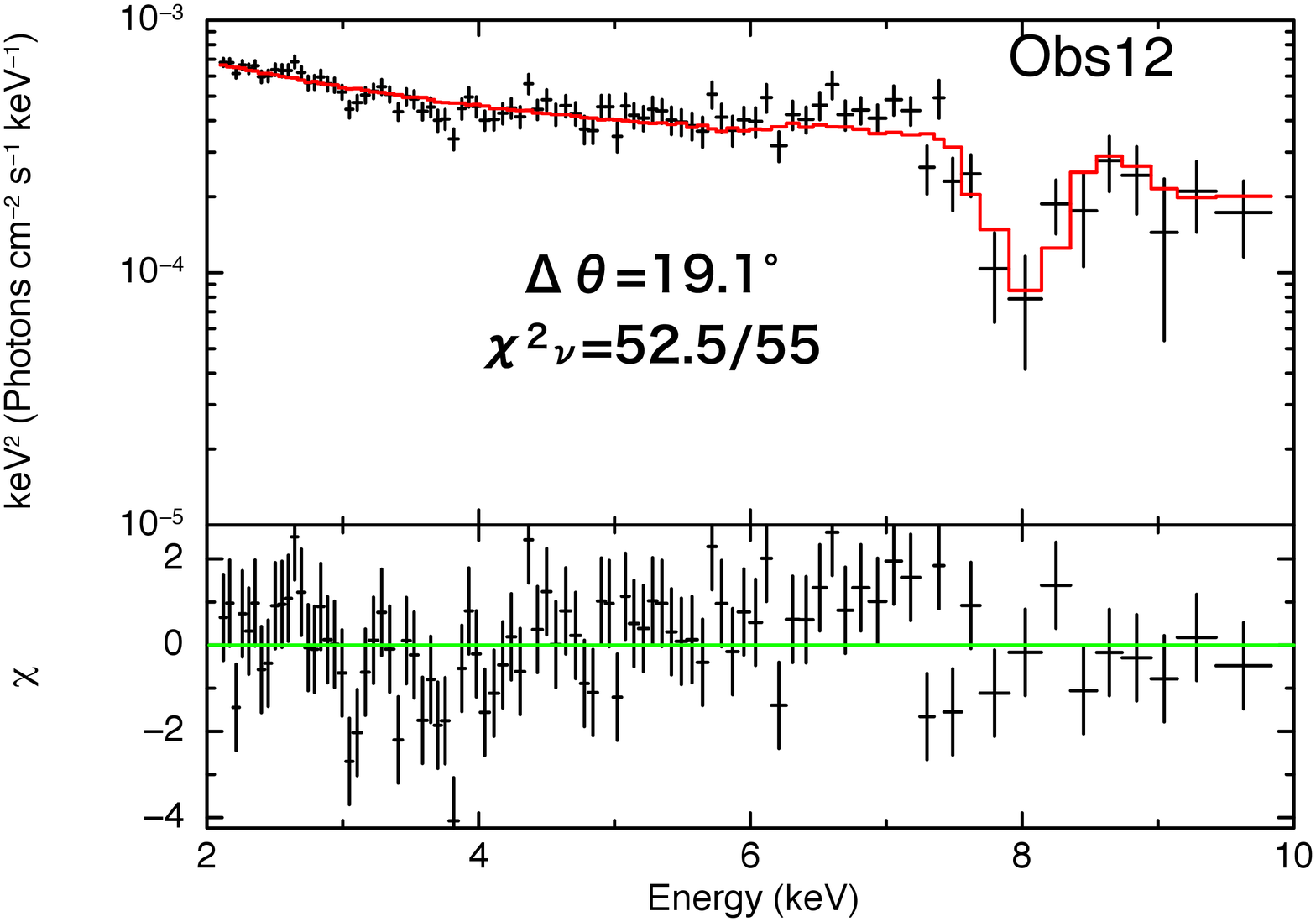}
\includegraphics[width=0.45\hsize]{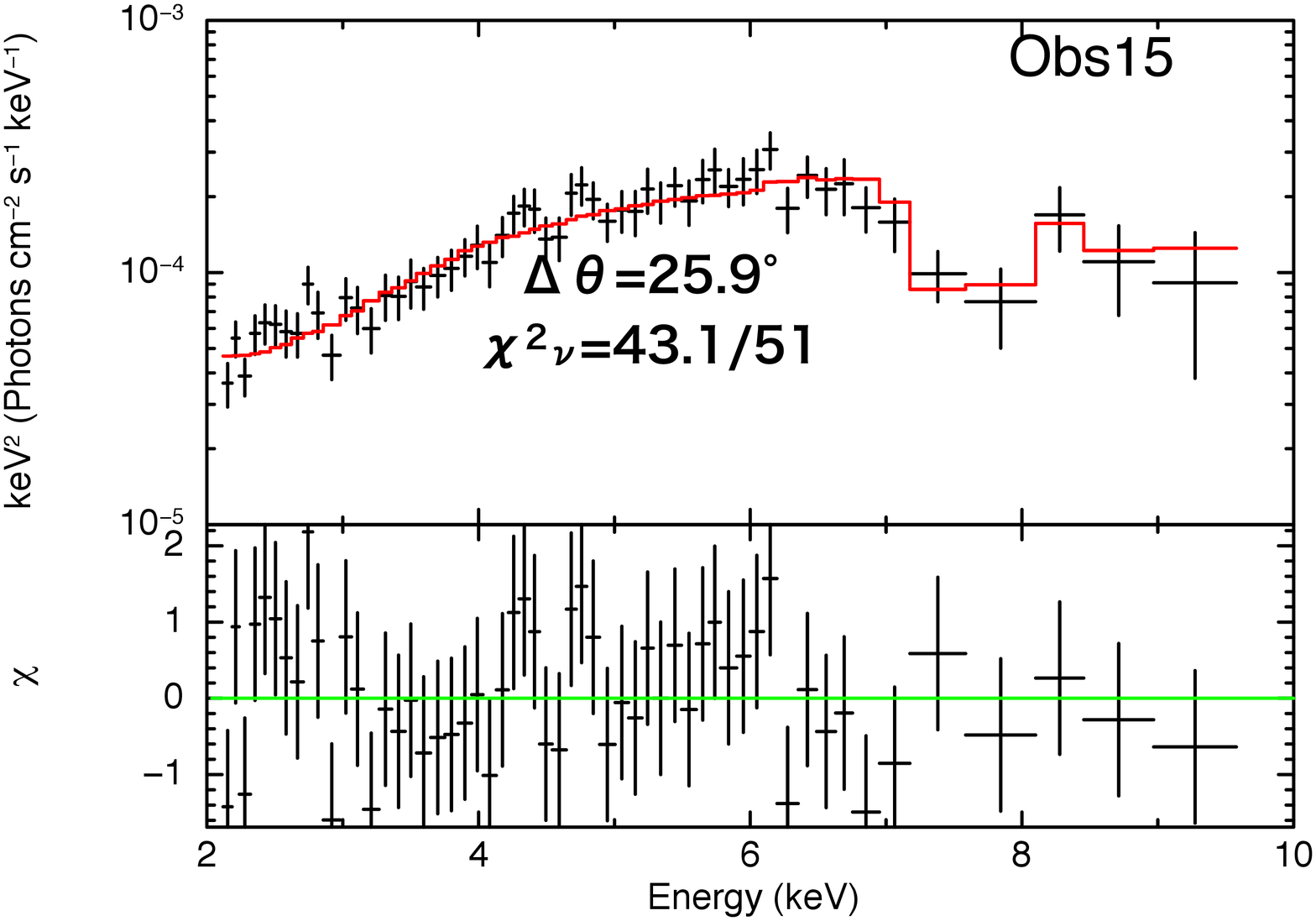}
\vspace{-0.2cm}
\caption{\textit{XMM-Newton} spectra of 1H~0707$-$495 compared with our spectral model. $\Delta\theta$ indicates a difference between our line of sight and the inner stream line of the wind \citep{Hagino2016}.}
\label{fig:1h}
\end{figure*}

We apply our disk wind model to all 15 spectra of 1H~0707$-$495 observed by \textit{XMM-Newton} and \textit{Suzaku}, and successfully reproduce the structure at iron-K band by changing only the viewing angle (Fig.~\ref{fig:1h}). It means that the spectra of this source can be explained by the UFO with $\dot{M}_\mathrm{wind}/\dot{M}_\mathrm{Edd}=0.2$ and $v_\mathrm{wind}=0.2c$. In addition to these data, we also compare an extrapolation of the best-fit wind model for Obs15 (right panel of Fig.~\ref{fig:1h}) with the \textit{NuSTAR} data \citep{Kara2015}. We find that the higher energy data is also explained by our disk wind model. This strongly supports our disk wind interpretation for this source.

\section{Conclusions}
We have constructed a new spectral model by calculating the radiation transfer in a realistic wind geometry based on the UV-line driving mechanism. We first applied our disk wind model to an archetypal wind source PDS~456, and found that the strong spectral variation in this source can be explained by a change of the viewing angle without changing a mass outflow rate. It indicates that the spectral variability is due to a local instability or inhomogeneity of the wind. As the next target, we applied to the ``disk-line'' source 1H~0707$-$495. Strong Fe-K features in all the spectra of this source observed by \textit{XMM-Newton}/\textit{Suzaku} are successfully reproduced by our disk wind model. Moreover, higher energy spectra obtained by \textit{NuSTAR} are also explained by our disk wind model.

\label{last}

\end{document}